\documentclass[prb,reprint,groupedaddress,showpacs]{revtex4-1}

\usepackage{graphicx}
\usepackage{dcolumn}
\usepackage{bm}

\begin{document}

\title{Nonuniform and coherent motion of superconducting vortices in the picometer-per-second regime}

\author{Jonghee Lee}
\email{jonghee@umd.edu}
\affiliation{Department of Physics, University of Maryland, College Park, Maryland 20742, USA,}
\affiliation{Laboratory for Physical Sciences, 8050 Greenmead Drive, College Park, Maryland 20740, USA,}
\affiliation{Department of Materials Science \& Engineering, University of Maryland, College Park, Maryland 20742, USA}

\author{Hui Wang}
\affiliation{Department of Physics, University of Maryland, College Park, Maryland 20742, USA}
\affiliation{Laboratory for Physical Sciences, 8050 Greenmead Drive, College Park, Maryland 20740, USA}

\author{Michael Dreyer}
\affiliation{Department of Physics, University of Maryland, College Park, Maryland 20742, USA}
\affiliation{Laboratory for Physical Sciences, 8050 Greenmead Drive, College Park, Maryland 20740, USA}

\author{Helmuth Berger}
\affiliation{Institut de Physique de la Mati\`{e}re Complexe, EPFL, CH-1015 Lausanne, Switzerland}%

\author{Barry I. Barker}
\affiliation{Laboratory for Physical Sciences, 8050 Greenmead Drive, College Park, Maryland 20740, USA}

\date{\today} 

\begin{abstract}
We investigated vortex dynamics in a single-crystal sample of type-II superconductor NbSe$_{2}$ using scanning tunneling microscopy at 4.2~K. The decay of the magnetic field at a few nT/s in our superconducting magnet induced the corresponding motion of vortices at a few pm/s. Starting with an initial magnetic field of 0.5~T, we continued to observe motion of vortices within a field of view of 400$\times$400~nm$^2$ subject to decay of the magnetic field over a week. Vortices moved collectively, and maintained triangular lattices due to strong vortex-vortex interactions during the motion. However, we observed two peculiar characteristics of vortex dynamics in this superconductor. First, the speed and direction of the vortex lattice motion were not uniform in time. Second, despite the non-uniform motion, we also found that there exists an energetically favored configuration of the moving vortices in the single-crystal sample of NbSe$_{2}$ based on the overlaid trajectories and their suppressed speeds. We model the system with weak bulk pinning, strong bulk pinning, and edge barrier effects.
\end{abstract}



\maketitle
Understanding vortex motion in superconductors plays a key role in developing useful superconductivity-based applications.\cite{csLee:vortex_ratchet,villegas:vortex_rectifier} Vortex matter\cite{bennemann:superconductivity,blatter:vortex-highTc} is also interesting in its own right, because a vortex system is a model for the studies of thermal and quantum fluctuations\cite{blatter:QM_melting}, self-organized criticality\cite{pak:soc,field:vortex_avalanche}, polymer physics\cite{nelson:melted_flux}, and Luttinger liquids\cite{kafri:Luttinger}. Although various techniques have yielded successes in imaging vortices and have become instrumental in our microscopic understanding of vortex dynamics\cite{harada:lorentz,moser:vortex-mfm,Pardo:bitter,troyanovski:vortex-motion,olsen:single-vortex-enter,guillamon:vortex_melting,kalisky:vortex_motion_at_grain_boundary}, it is still challenging to capture the motion of individual vortices when the inter-vortex spacing is less than 100~nm. To image vortices on this length scale or beyond, a scanning tunneling microscope (STM) is the right tool to be used because of the excellent spatial resolution. We are particularly interested in directly observing the response of vortices to an external driving force. Previously, direct observations of flux creep\cite{troyanovski:vortex-motion} without an external driving force and vortex lattice melting\cite{guillamon:vortex_melting} due to thermal excitations were reported. However, driving and observing vortex motion at predictable speeds using an external driving source has not been reported. Here we report on vortex motion on and off experiments on a single-crystal sample of type-II superconductor NbSe$_{2}$ by turning on and turning off a slow decay of the magnetic field. Our prediction of the averaged speed of the magnetically driven vortex motions agrees well with the magnetic field decay rate.

\begin{figure}
\begin{center}
\includegraphics[width=1.0\linewidth]{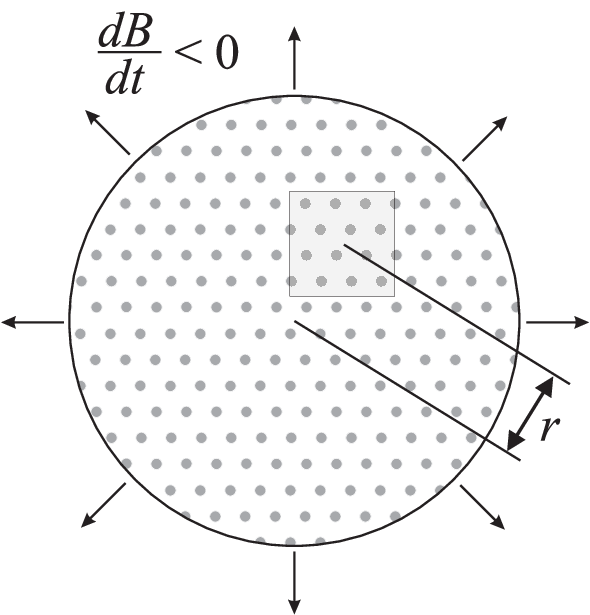}
\end{center}
\caption{\label{fig:vortex-motion-model} 
Vortex motion model subject to the decay of a magnetic field. As a magnetic field $B$ decays, vortices leave the sample. An observation within a field of view (rectangular area) is made at a distance $r$ away from the center of the sample. The size of the field of view, vortices, and $r$ are drawn for illustration, not to scale.}
\end{figure}

\begin{figure*}
\begin{center}
\includegraphics[width=1.0\linewidth]{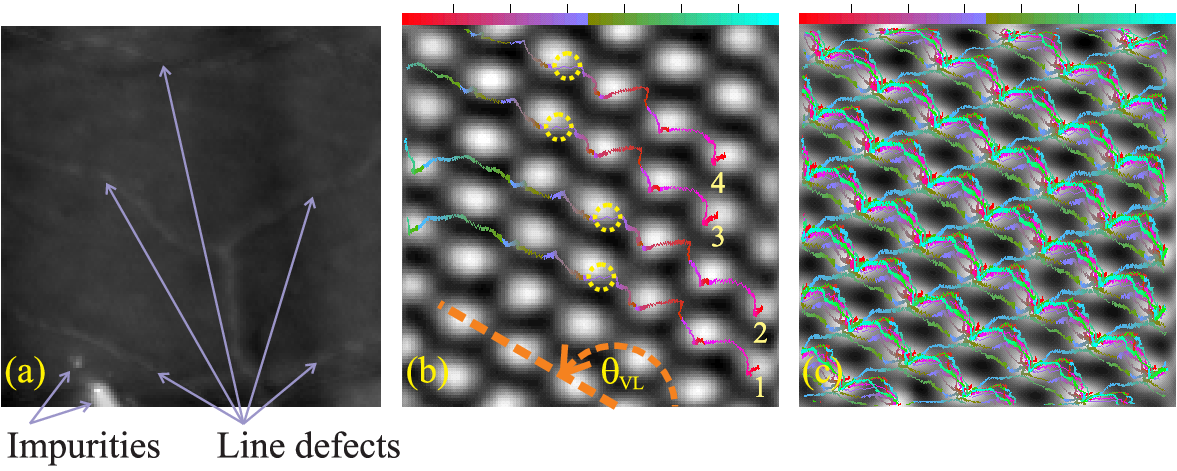}
\end{center}
\caption{\label{fig:vortex-motion} (Color online) (a) Averaged topography of NbSe$_2$ in the area of $400 \times 400$~nm$^2$, where consecutive vortex images were recorded during continuous scanning over 7~days. Signal-to-noise ratio of a single topographic image was not sufficient to reveal surface features such as line defects and impurities. Such features in the area are clearly revealed after averaging over 2560 topographic images. (b) Collective and coherent motion of an ordered vortex lattice. The underlying gray-scaled image, the first spectroscopic image of the 7-day observation, shows a snapshot of vortices moving at $\sim$pm/s within the field of view. Four time-color-coded trajectories of vortices 1, 2, 3, and 4 imply that vortices moved coherently as a whole. Progression of time is represented by the color bar with tick marks every 24 h. The yellow dotted circles emphasize the fastest motion as indicated by the blue arrows in Figs.~\ref{fig:vortex-speed}(a) and \ref{fig:vortex-speed}(b). The $\theta_{\mathrm{VL}}$ denotes the angle of one of the principle axes of the VL. (c) Unique configuration of the favored locations (bright sites in the grayscale image) for the minimum of total potential energy of the moving vortex system is clearly revealed after all trajectories (time color coded) are overlaid. The underlying gray-scaled image is the average of all 2560 vortex images. We used the following scanning and lock-in parameters: tunneling current $I_{\mathrm{t}}=0.1$~nA, bias voltage $V_{\mathrm{bias}}=3$~mV, scan speed $v_{\mathrm{scan}}=600$~nm/s, and \textit{ac} modulation voltage $\Delta V_{\mathrm{bias}}=1$~mV$_{\mathrm{rms}}$ at a modulation frequency $f = 1973$~Hz.}
\end{figure*}

When an external magnetic field between the lower and upper critical values is applied to a type-II superconductor below its critical temperature, magnetic vortices arrange themselves in the form of a triangular lattice, called a vortex lattice (VL), with a lattice constant\cite{tinkham:super}
\begin{equation}
\label{eq:lattice-const}
a(B)=\left(\frac{2}{\sqrt{3}}\right)^{1/2} \left(\frac{\Phi_{0}}{B}\right)^{1/2}~,
\end{equation}
where $B$ is the magnetic field penetrating the superconductor, and $\Phi_0$~($= 2.07 \times 10^{-15}~\mathrm{Tm}^2$) is the magnetic flux quantum. To drive vortex motion, we use a decaying magnetic field. Our superconducting magnet has an average decay rate of $\Delta B/\Delta t \approx - 4.2$~nT/s ($\approx - 0.36$~mT/day). This decay causes the triangular VL to expand (Fig.~\ref{fig:vortex-motion-model}), satisfying Eq.~(\ref{eq:lattice-const}). Using this decay rate and Eq.~(\ref{eq:lattice-const}), one can calculate the speed of vortex motion
\begin{equation}\label{eq:v-theory}
v_{\mathrm{theory}}=\frac{r}{2B}\cdot\left|\frac{dB}{dt}\right| 
\end{equation}
at a distance $r$ away from the center of the sample as shown in Fig.~\ref{fig:vortex-motion-model}. Since we used a roughly round disk of single-crystal NbSe$_{2}$ with a diameter of 5~mm (thickness = 0.5~mm) and positioned the scanning tunneling microscope (STM) tip near the center of the sample, we expected uniform motion, not exceeding $\sim$10~pm/s.

To observe such a slow motion we used a home-built STM system\cite{dreyer:lps-lt-stm} operating at liquid-helium temperature (4.2~K). The sample was cleaved at room temperature under a pressure of $\sim$6$\times$10$^{-7}$~mbar and transferred to the STM. After thermal equilibrium was reached, the initial magnetic field was set to 0.500~T. Next, we focused on vortex motion in a fixed area of 400$\times$400~nm$^2$ and continuously scanned down and up while recording consecutive vortex images in the area over 17~days. Using automated data analysis we extracted the time-series of speed $v(t)$ for all the vortices\cite{misc:vortex_data_analysis}. In this Rapid Communication, we present and discuss specifically the data taken during the last 7 days, ignoring earlier data for two reasons. One is to avoid as much as possible any transient effects\cite{yamamoto:field_decay} after the initial ramp-up of the magnetic field. The other is that this was the longest continuous observation without any interruption of data collection, such as refilling the liquid-helium dewar. This 7-day data series contains 2560 images of VL, for a total of 102810 vortex data points.

\begin{figure}
\begin{center}
\includegraphics[width=1.0\linewidth]{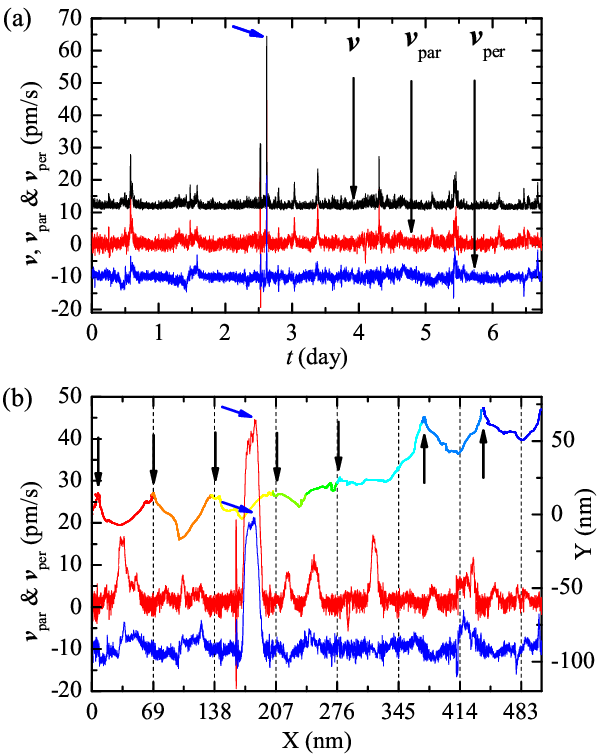}
\end{center}
\caption{\label{fig:vortex-speed} (Color online) Speeds of the VL motion. (a) Speeds ($v$, $v_{\mathrm{par}}$, and $v_{\mathrm{per}}$) vs. time ($t$). $v_{\mathrm{par}}$ and $v_{\mathrm{per}}$ are the components of the speed $v$ of the VL, parallel and perpendicular to one of the principal axes of the VL ($\theta_{\mathrm{VL}} = 150^\circ$). The data of $v$, $v_{\mathrm{par}}$, and $v_{\mathrm{per}}$ are smoothed by averaging over nine points to reduce noise. $v$ and $v_{\mathrm{per}}$ are shifted vertically by $\pm$10~pm/s, respectively, for clarity. (b) Displacements and speeds. The displacements [$X(t), Y(t)$] of the VL can be expressed as $\int^{t}_{0} v_{\mathrm{par}}(t)dt$ and $\int^{t}_{0} v_{\mathrm{per}}(t)dt$, respectively. The plot of [$X(t), Y(t)$] is discretely color coded every 23 h. The down and up arrows indicate the moments when vortices are at the favored locations [bright sites in Fig.~\ref{fig:vortex-motion}(c)]. The dotted grid lines are drawn at every 69~nm, corresponding to the lattice constant of the VL at 0.496~T. The red and blue curves are the parametric plots of [$\mathrm{X}(t), v_{\mathrm{par}}(t)$] and [$\mathrm{X}(t), v_{\mathrm{per}}(t)$] respectively, where [$\mathrm{X}(t), v_{\mathrm{per}}(t)$] is shifted vertically by $-10$~pm/s for clarity. The blue arrows in (a) and (b) indicate the fastest motion which occurred within the yellow dotted circles in Fig.~\ref{fig:vortex-motion}(b).}
\end{figure}

\begin{figure}
\begin{center}
\includegraphics[width=1.0\linewidth]{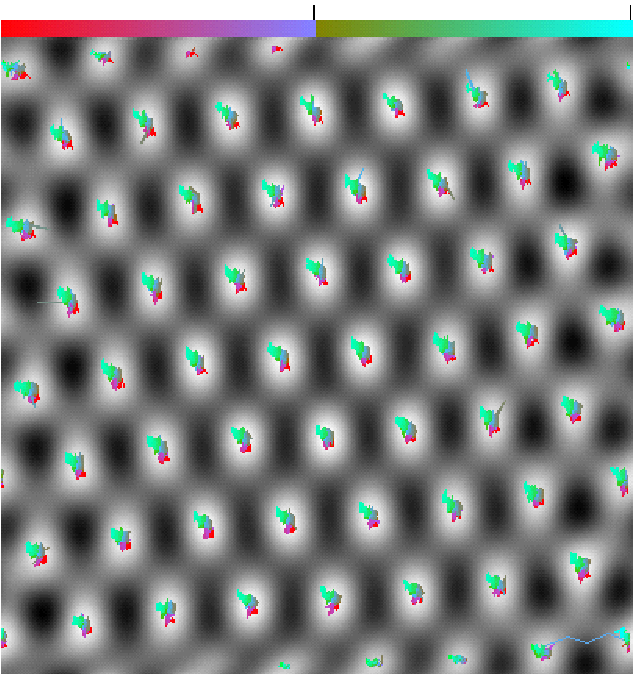}
\end{center}
\caption{\label{fig:vortex-stop} (Color online) Turning off vortex motion. One can clearly see stationary vortices based on the overlaid trajectories (time color) over 2 days. The measurement time was significantly reduced because the liquid-helium boil-off rate was increased with the magnet leads submerged in liquid helium in order to continuously supply current during the measurement. The gray-scaled image is averaged over 704 VL images. The observation was made in an area of $400 \times 400~\mathrm{nm}^2$ under 0.75~T. We used $I_{\mathrm{t}} = 0.1~\mathrm{nA}$, $V_{\mathrm{bias}} = 3~\mathrm{mV}$, $v_{\mathrm{scan}}=551$~nm/s, and $\Delta V_{\mathrm{bias}}=1$~mV$_{\mathrm{rms}}$ at $f = 1973$~Hz.}
\end{figure}

Figure~\ref{fig:vortex-motion}(a) shows the averaged topographic image in the field of view. One can clearly see several line defects and impurities on the surface. The gray-scaled image in Fig.~\ref{fig:vortex-motion}(b) is the first spectroscopic image in the 7-day data series, showing a regular triangular VL with a lattice constant of 69~nm. As the magnetic field decayed, the VL moved mostly along one of its own principal axes ($\theta_{\mathrm{VL}}=150^{\circ}$), while maintaining its ordered structure\cite{vortex_movie}. As the time-colored trajectories of four vortices (labeled as 1, 2, 3, and 4) indicate, the motion of the VL was non-uniform in terms of speed and direction. Despite the non-uniform motion, the inter-vortex spacing remained almost constant throughout the measurement. Even at the highest speed, which is 45 times faster than the averaged speed [marked as the yellow dotted circles in Fig.~\ref{fig:vortex-motion}(b) and the blue arrows in Figs.~\ref{fig:vortex-speed}(a) and \ref{fig:vortex-speed}(b)], the ordered structure of the VL was still maintained. This rigidity of the VL indicates that a strong vortex-vortex interaction dominates distortions of the VL due to local pinning forces. After analyzing the 2560 images of the VL, we found that the lattice constant expanded by $\sim$0.3~nm over 7 days in comparison to $\sim$0.2~nm, as would be expected from our known field decay rate. However, we consider that this amount of change is sufficiently small compared to the initial lattice constant of 69~nm, so that one can assume that a single vortex motion represents the dynamics of the whole VL to a good approximation. This allowed us to combine all the time series of individual vortices into a single time series by sorting the observed speeds by time [black in Fig.~\ref{fig:vortex-speed}(a)]. Then we decomposed $v$ into $v_{\mathrm{par}}$ (red) and $v_{\mathrm{per}}$ (blue) ($v_{\mathrm{par}}$ and $v_{\mathrm{per}}$ are the speeds along and perpendicular to the direction of $\theta_{\mathrm{VL}}$, respectively.). We found that $\langle v \rangle$, $\langle v_{\mathrm{par}}\rangle$, and $\langle v_{\mathrm{per}}\rangle$ are 2.5, 0.9, and 0.1~pm/s, respectively. The observed motion at $\sim$pm/s agrees well with the estimated speed according to Eq.~(\ref{eq:v-theory}). However, to better understand the non-uniform motion of VL in a single-crystal superconductor, it is necessary to include some pinning effects which may originate either from bulk or from the edges of the sample. We will discuss these later.

Strikingly, by overlaying the trajectories of all the vortices over 7 days [color in Fig.~\ref{fig:vortex-motion}(c)], we found that vortices repeatedly visited favored locations more often than other regions, although vortices arrived at the favored locations along different paths. This characteristic resulted in the 2560-averaged VL image (gray-scaled), which represents the distribution of the probability of finding a vortex at a certain location. The bright sites, where vortices are more likely to be found than in dark regions, clearly form a triangular lattice with its lattice constant close to that of the VL [Fig.~\ref{fig:vortex-motion}(b)].

Another way of looking at our vortex dynamics is to see temporal and spatial characteristic by using a variable transformation $X = \int^{t}_{0} v_{\mathrm{par}}(t)dt$ and $Y = \int^{t}_{0} v_{\mathrm{per}}(t)dt$. We refer to $(X,Y)$ as a displacement of VL. Using this transformation, all trajectories collapse into one single trace, as shown in Fig.~\ref{fig:vortex-speed}(b). Therefore, one can conveniently see the spatial variation of speeds as the VL moves along the single trace of $(X,Y)$ displacement. The down and up arrows mark the moments when vortices are at the favored locations. When vortices are at the favored locations, both $v_{\mathrm{par}}$ and $v_{\mathrm{per}}$ are suppressed such that $\sqrt{[v_{\mathrm{par}}(t)]^{2}+[v_{\mathrm{per}}(t)]^{2}}< 3.5$~pm/s. However, while vortices are transiting from favored locations to other favored locations, accelerations and decelerations occur, showing fast motion such that $\sqrt{[v_{\mathrm{par}}(t)]^{2}+[v_{\mathrm{per}}(t)]^{2}} > 5$~pm/s. Therefore, the high probability and the suppressed speeds at the favored locations indicate that the favored locations make up the configuration for the minimum of total potential energy of the moving vortex system. That configuration was preserved over 7 days.

To confirm that the vortex motion was caused mainly by the decay of the magnetic field, we attempted to stop the vortex motion by stabilizing an initial magnetic field. The initial magnetic field was stabilized by continuously supplying the current to the magnet during measurement. First we ramped up the magnetic field from 0 to 0.75~T, stabilized it, and immediately started the measurement. Then we continued to monitor vortex motion over 2 days. In Fig.~\ref{fig:vortex-stop}, one can clearly see a quite reduced motion of vortices based on the overlaid trajectories. The $\langle v_{\mathrm{par}} \rangle$ and $\langle v_{\mathrm{par}} \rangle$ were reduced by a factor of 10 compared to those of the moving VL. Therefore, we confirmed that the decay of the magnetic field was the major driving source for the vortex motion shown previously, and we demonstrated that vortex motion can be turned on or off depending on the method of operation of a superconducting magnet.

Although we experimentally verified that our vortex motion was driven mainly by the decay of the magnetic field, there are still two open questions in our observation. One is, what causes the non-uniform motion in terms of direction [Figs.~\ref{fig:vortex-motion}(b) and \ref{fig:vortex-motion}(c)] and speed [Figs.~\ref{fig:vortex-speed}(a) and \ref{fig:vortex-speed}(b)]? The other is, despite the non-uniform motion, why do the favored locations exist as shown in Fig.~\ref{fig:vortex-motion}(c)? Currently we are not able to answer these two questions. We discuss possible causes of these features below.

We divide possible causes of the non-uniform motion in a single-crystal superconductor into two categories. One is bulk pinning and the other is the edge effects. As for the bulk pinning, randomly distributed weak pinning sites\cite{larkin:cp} within the single-crystal sample may become a dominant cause of the non-uniform vortex dynamics. For example, even without a driving source, one may observe vortex motion in the flux creep regime at a constant magnetic field because the motion of vortices can be impeded due to weak pinning sites. Indeed Troyanovski~\textit{et al.}\cite{troyanovski:vortex-motion} directly observed this motion in a single-crystal NbSe$_2$ by using a STM. They observed flux creep motion on the order of nm/s in 0.6~T, 20 min after the initial magnetic field was established. However, we did not observe such a fast creep motion in 0.75~T as shown previously. This comparison may indicate that our single-crystal sample has a much weaker ``weak pinning'' nature than their single-crystal sample does. Nevertheless, we do not rule out the possibility that such weak pinning sites may contribute to the non-uniform motion because our vortex system was driven at such slow speeds of pm/s.

The other possibility is that the non-uniform motion may be affected by some effects coming from the edges of the sample, such as Bean-Livingston surface barriers\cite{Bean:surface-barrier} or edge pinnings\cite{kouzoudis:edge_barrier_pinning} due to imperfections along the sample edges of a real material. In an ideal superconductor without any defect but with a boundary between the superconductor and vacuum space, according to the Bean-Livingston surface barrier model, there exists a potential energy barrier on the boundary that prevents a vortex from entering and exiting the superconductor\cite{Bean:surface-barrier}. This potential energy barrier originates from the competition between the repulsion from the boundary caused by the external magnetic field penetrating into the superconductor, and the attraction to the boundary that is explained by the interaction between the vortex and the mirror-imaged vortex. Therefore, for a vortex to enter or exit the superconductor, the vortex needs to overcome this potential barrier on the boundary, called the Bean-Livingston surface barrier. Moreover, in a real sample, imperfections along the edges of a real material can also prevent vortices from entering and exiting the superconductor\cite{kouzoudis:edge_barrier_pinning}.  As a result, the vortex crossing the edges of the superconductor is not smooth but sudden. Indeed, this phenomenon can be directly observed by the magneto-optical method\cite{olsen:single-vortex-enter} in a relatively weak magnetic field. These vortex entering and exiting events disturb the arrangement of vortices on the vicinity of the edges. This disturbance on the edges may be coupled to our non-uniform vortex motion through strong vortex-vortex interactions.

To check how boundaries of a sample affect the vortex dynamics, we carried out a two-dimensional molecular simulation study using a large number of vortices (7729 initially) in a rectangular area of 8$\times$4~$\mu$m$^2$ (Ref.~\onlinecite{dreyer:vortex_simulation}). First we did not include any bulk pinning, but we assumed repulsive exponential forces exerted on the vortices from the boundaries when the vortices are close to the boundaries of the area. We employed repulsive forces in the form of a Bessel function between any two vortices. Then, to drive the vortex motion, we extracted vortices one at a time at a fixed location on one of the boundaries. The time intervals between the vortex extractions were randomly chosen between 10 and 30~s. In a real situation, vortices exit the sample randomly in time, in space, and  in the number of exiting vortices. We, however, impose randomness only on the time interval, for simplicity.

We found that vortex lattice dislocations were formed due to stress buildup, and dislocation lines of the vortex lattices propagated through the area as vortices were extracted (see Fig.~11 in Ref.~\onlinecite{dreyer:vortex_simulation}). As a result this caused random spikes in the speed variation of vortex motion even without any bulk pinning sites (see Fig.~10 in Ref.~\onlinecite{dreyer:vortex_simulation}). Even with some defects introduced into the area, we found that their contribution to non-uniform vortex motions was small compared to that due to propagation of the vortex lattice dislocation lines triggered by strong vortex-vortex interactions and boundary conditions as vortices were extracted from the area. In our experiment, however, we did not observe any propagation of vortex lattice dislocation lines within the field of view. This may be because our field of view in the experiment was not large enough to cover the widths of dislocation lines. If a dislocation line has a certain width, and the width is larger than the size of the field of view, one cannot see the dislocation line within the small field of view, but one can still observe the sudden changes of VL speed, when the dislocation line passes the field of view.

We do not know which one is the more dominant cause of the non-uniform motion: bulk pinning or edge effects. In order to experimentally discriminate edge effects from bulk pinning, one needs to directly observe vortex motion in a Corbino-geometry disk sample\cite{Paltiel:corbino}. In a Corbino-geometry disk, currents flow radially between the center and perimeter of the disk, and the force acting on each vortex due to currents is perpendicular to the radial direction. Therefore, vortices circulate within the sample, not crossing the sample edges. Thus one can extract the effects caused by bulk pinning by minimizing the edge effects.

The most intriguing question is why vortices energetically favor certain locations despite the non-uniform motion. We study this aspect using our 2D molecular dynamics simulation\cite{dreyer:vortex_simulation} by including strong and weak bulk pinning and surface barrier effects into the model area of $8\times4$~$\mu$m$^2$. As for the strong pinning case, the triangular lattice regularity of the moving VL breaks around a strong pinning site. Therefore, one would not expect to see the existence of favored locations as experimentally observed. Away from the strong pinning sites, moving vortices restore the triangular lattice regularity because of strong vortex-vortex interactions. However, it does not show the favored locations. As for the weak pinning case, the triangular lattice regularity is sufficiently maintained around a weak pinning site, and vortices smoothly pass by the weak pinning site, resulting in almost straight trajectories. We also do not observe favored locations in this case. However, it would be possible that we do not observe the favored locations in our simulation because the number of vortices (7729) used in the simulation is still not sufficient. We are currently limited by our computing capacity, and a more extensive simulation is left for future research.

In summary, we show the direct observation of vortex dynamics in a single-crystal NbSe$_2$ sample in the pm/s moving regime. We observe peculiar characteristics of vortex dynamics: non-uniform motion and the existence of an energetically favored configuration of vortices. Although we are not able to definitively identify the cause(s) of the two characteristics of our observation, we suspect the non-uniform motion and the existence of favored locations may arise from the pinning sites in bulk or the edge effects along the sample edges or a combination of the two. The causes of the existence of a unique configuration of favored locations in the single-crystal sample of NbSe$_2$ are currently unclear. For future experimental research, to distinguish bulk pinning effects from edge effects, one may attempt to observe vortex motion in a Corbino-geometry disk sample\cite{Paltiel:corbino} by using a STM.  In addition, application of our vortex driving method to pinning engineered\cite{karapetrov:geometrical-phase-transition} or \emph{in situ} defect-induced\cite{hui:cdw} NbSe$_2$ samples, and comparison of those vortex motions with the motion in single-crystal NbSe$_2$ reported in this Rapid Communication, would be helpful in better understanding the vortex dynamics in the picometer-per-second regime.

We acknowledge Douglas Osheroff, Christopher Lobb, Danilo Romero, Ichiro Takeuchi, Young-noh Yoon, and Late Sung-Ik Lee for valuable discussions.


\begin{thebibliography}{28}
\bibitem [{mis()}]{misc:vortex_data_analysis}%
  \BibitemOpen
  \href@noop {} {}\bibinfo {note} {For each vortex, $\sim$77 pixels were used
  to determine its center position by calculating the ``center of mass.'' Each
  ``mass'' corresponds to the scaled value at a given pixel. This improved the
  spatial resolution from 3.125 to 0.356 nm ($= 3.125/\sqrt{77}$). We also
  calculated the time at which the center of a vortex was reached by the STM
  tip, using the scan speed. The speed $v$ of a vortex in the $n$th frame was
  calculated by the travel distance and the detected time difference of the
  vortex between $(n-1)$th and $(n+1)$th frames. This method eliminated the
  systematic errors which occur when calculating $v$ using two adjacent frames
  [$n$th and $(n+1)$th] due to hysteresis of the STM scanner. We achieved the
  $v$ resolution of 0.74~pm/s [$=0.356~\mathrm{nm}/(2\times 4~\mathrm{min})$]
  in a single measurement. This limit was further improved by averaging more
  data points and continuous observation.}\BibitemShut {Stop}%
\end{thebibliography}
\end{document}